\documentclass[12pt,a4paper]{article}
\usepackage[T1]{fontenc}
\usepackage[english]{babel}
\usepackage{graphics}
\usepackage{setspace}

\input tcilatex
\begin{document}

\title{Comparative Monte Carlo Study of a Monolayer Growth in a Heteroepitaxial
System in the Presence of Surface Defects}
\author{M. Cecilia Gim\'{e}nez, Ezequiel P. M. Leiva \thanks{%
Corresponding author. Fax 54-351-4344972; e-mail: eleiva@fcq.unc.edu.ar} \\
Unidad de Matem\'{a}tica y F\'{\i}sica, Facultad de Ciencias Qu\'{\i}micas \\
INFIQC.\\
Universidad Nacional de C\'{o}rdoba, 5000\\
C\'{o}rdoba, Argentina}
\maketitle

\begin{abstract}

The adsorption of a metal monolayer or submonolayer for the
systems Ag/Au(100), Au/Ag(100), Ag/Pt(100), Pt/Ag(100), Au/Pt(100),
Pt/Au(100), Au/Pd(100) and Pd/Au(100) was studied 
by means of lattice Monte Carlo simulations.

It was found that, taking into account some general trends, such systems can be
classified into two big groups.
The first one comprises 
Ag/Au(100), Ag/Pt(100), Au/Pt(100) and Au/Pd(100), which have
favourable binding energies as compared with the homoepitaxial growth of
adsorbate-type atoms. When the simulations are performed in the presence of
substrate-type island in order to emulate surface defects, the islands
remain almost unchanged, and the adsorbate atoms successively occupy kink sites, 
step sites and the complete monolayer.

The second group is composed of Au/Ag(100), Pt/Ag(100),
Pt/Au(100), and Pd/Au(100), for which monolayer adsorption is
more favourable on substrates of the same nature than on the considered
substrates.
When simulations are carried out in the presence of islands of 
substrate-type atoms,
these tend to disintegrate in order to form 2-D alloys with adsorbate atoms.

\end{abstract}

\section{Introduction}

Research on the electrochemical deposition of a metal $M$ onto a well-ordered
single crystalline surface of a foreign metal $S$ should provide a better
understanding of the fundamental aspects of metal deposition [1-8].
When this occurs at potentials more
positive than those predicted from the Nernst equation, the process is
denominated underpotential deposition(\textbf{upd}) \cite{Kolb, Trasatti,
Blum, Lorenz_2}, to differenciate it from the usual deposition
process at overpotentials. \ Due to the high complexity of the deposition
phenomenon, no general theory has been able to embody the behavior
of metal monolayers and submonolayers formed on a foreign single-crystal 
surface. This is partly due to the fact that, even in this 
simple case, the process is complicated by the appearance of several
phases, involving expanded structures, anion adsorption, formation of
surface alloys, etc. 
While each system may require its own  specific modelling in order for 
its fine features to be properly described, we think that comparative 
computational studies of some simple model systems may yield important 
information leading to understanding of the more complicated experimental 
systems.
Thus, rather than seeking ''experimental agreement'', 
it may be relevant to analyze a model that contains some key features 
that play an important role in the experiments.

It is the purpose of this work to analyze the growth of a metal on a foreign
substrate in the presence of surface defects that will be represented by
islands of the same material as that of the substrate. 
This is done in order to find trends or regularities for a
number of systems involving metals of common use in electrochemistry. 
As they provide a realistic interaction of the metallic binding, EAM potentials
\cite{Daw-Baskes} are employed to describe the atom-atom interaction.
Using the Monte Carlo method within a lattice model, allows us to deal with
systems having a reasonably large number of particles.

Although the model chosen may seem simple as compared with reality,
the rich behavior of the present simulations provides several
clues to understand the behavior of some experimental systems, and suggests a
number of future studies that may be tackled by experimentalists in the
area. The present comparative study involves the systems Ag/Au(100),
Au/Ag(100), Ag/Pt(100), Pt/Ag(100), Au/Pt(100) and Pt/Au(100),Pd/Au(100)
and Au/Pd(100).

\section{Model and simulation method}

\subsection{Lattice model}

As they allow to deal with a large number of particles at a relatively 
low computational cost, lattice models are widely used in computer simulations
to study nucleation and growth.
In principle, it must be
kept in mind that continuum Hamiltonians, where particles are allowed to
take any position in space, are much more realistic in cases where
epitaxial growth of an adsorbate leads to incommensurate adsorbed phases 
\cite{Pb/Ag} or to adsorbates with large coincidence cells. 
On the other hand, to assume that particle adsorption can only occur at
definite sites is a good approximation for some systems.
Such is the case of silver
on gold, where there is no crystallographic misfit. We have studied this
system before \cite{Ag-Au-thermo} and now we apply the same assumption to
other systems in order to perform a comparative analysis when adsorption
takes place in the presence of surface defects.

With this purpose, we employ here a lattice model to represent the square
(100) surface lattice in a Grand Canonical Monte Carlo simulation. Besides
adsorbate adatoms, atoms of the same nature as the substrate are
present in the monolayer at a coverage degree of 0.1, with different
surface structures so as to emulate some of the most common
surface defects. 
In order to obtain the adsorption isotherms and study the influence of the
surface defects on their shapes, we calculate the average coverage degree
at each chemical potential. 
Solvent effects are neglected, but this
should not be a major problem for the metal couples considered, since the
partial charge on the adatoms is expected to be small, thus minimizing the
dipole-dipole interactions. We also neglect all kinds of anion
effects that may coadsorb during the metal deposition process. This may lead
to some underestimation or overestimation of the underpotential shift
defined below \cite{Kolb} depending on wether anions adsorb more
strongly or more weakly on the adsorbate than on the substrate\cite
{first-p-upd}.

Square lattices of size $(100\times 100)$ with periodical boundary
conditions are used here to represent the surface. Each lattice node
represents an adsorption site for an atom. The adsorbate may adsorb, desorb
or hop between neighboring sites, while the atoms of the same nature as the
substrate may only move on the surface like the other atoms do. In this way,
our model corresponds to an open system for one of its components, as it is
the case of adatom deposition on a foreign surface.

Different structures can be chosen as initial conditions for each
simulation. In the present case, we start with islands of atoms of the same
nature as the substrate of different sizes and shapes obtained by means of
simulated annealing techniques. This was undertaken in order to imitate some
of the defects that can be found on a real single crystal surface, like kink
sites, isolated substrate atoms, steps, etc.
At each chemical potential, the initial condition is the monolayer with
surface defects generated by simulatted annealing of 1000 substrate atoms, 
as will be discussed later, and 2000 adsorbate atoms, giving an initial coverage
degree of $\theta = 0.222$.

\subsection{Energy Calculation}

To calculate the activation energies for adatom diffusion the Embedded Atom
Method (EAM) was used \cite{Daw-Baskes}. This method takes into account 
many-body effects; therefore, it represents the metallic bonding better than a
pair potential does. The total energy of the system is calculated as the sum
of the energies of the individual particles. Each energy is in turn the sum
of an embedding (attractive) energy and a repulsive contribution which arises
from the interaction between ion cores. The EAM contains parameters which
were fitted to reproduce experimental data, such as elastic constants,
enthalpies of binary alloys dissolution, lattice constants, and sublimation
heats.

The application of the EAM within a lattice model is described in detail in
our previous work \cite{Ag-Au-thermo}. In the present work some numerical
results may differ from those obtained in the previous one due to the fact 
that we had used
'single zeta' functions and now we use 'double zeta' functions for the
calculation of the energies \cite{tablas_eam_1, tablas_eam_2}. Nevertheless, in
both works the qualitative results and the conclusions are the same.

\subsection{Grand Canonical Monte Carlo}

One of the most appealing characteristics of Grand Canonical Monte Carlo $%
(\mu VT/MC)$ is that, like in many experimental situations,
the chemical potential $\mu $ is one of the independent variables.
This is the case of low-sweep rate voltammetry,
an electrochemical technique where the
electrode potential can be used to control the chemical potential of species
at the metal/solution interface. This technique offers a straightforward way
of obtaining the adsorption isotherms provided the sweep rate is low enough
to ensure equilibrium for the particular system considered. In the
solid-vacuum interface, the chemical potential is related to the vapor
pressure of the gas in equilibrium with the surface. However, chemical
potential sweeps under equilibrium conditions cannot be achieved in the
solid-vacuum interface because the desorption process is too slow due to the
high energy barrier that the metal adatoms have to surmount (typically of
the order of some eVs).

Our 2D system is characterized by a square lattice with $M$ adsorption
sites. We labelled each adsorption site $0$, $1$ or $2$,
depending on whether it is empty, occupied by one substrate type atom or
occupied by one adsorbate atom respectively.

Following the procedure proposed by Metropolis and coworkers \cite{Allen},
the acceptance probability for a transition from state $\overrightarrow{n}$
to $\overrightarrow{n}^{\prime }$ is defined as:

\begin{equation}  \label{8}
W_{\overrightarrow{n}\rightarrow \overrightarrow{n}^{\prime}}=\min (1,\frac{%
P_{\overrightarrow{n}^{\prime}}}{P_{\overrightarrow{n}}})
\end{equation}

{\raggedright so that detailed balance is granted.}

In our $\mu VT\ /\ MC$ simulation we shall allow for three types of events:

\begin{enumerate}
\item  Adsorption of an adsorbate atom onto a randomly-selected lattice site.

\item  Desorption of an adsorbate atom from an occupied lattice
site selected at random.

\item  Motion of an atom from the lattice site where it is adsorbed to one
of its four nearest neighbor sites. The latter is selected randomly.
\end{enumerate}

Even when in a grand canonical simulation, events of type 3 (motion or
diffusion of lattice particles) are not strictly necessary, their presence
is justified by the fact that a smaller number of Monte Carlo steps(MCS)
must be employed for the equilibration of the system \cite{Allen}. It is also
important in the case of motion of substrate type atoms, which would not be
possible without that type of events.

Within this procedure, the relevant thermodynamic properties are then
obtained, after some equilibration steps, as average values of 
instantaneous magnitudes stored along a simulation run. 
In the present case, we are
interested in the average coverage degree of the adsorbate atoms $%
\left\langle \Theta \right\rangle _{Ads}$ at a given chemical potential $\mu$, 
where the instantaneous value, $\Theta (\mu)_{Ads,i}$ , is
defined as follows:

\begin{equation}
\Theta (\mu)_{Ads,i}=\frac{N_{Ads,i}}{M-N_{Su}}  \label{13}
\end{equation}

\noindent where $N_{Ads,i}$ is the number of adsorbate atoms, 
$M$ is the total number of sites and $N_{Su}$ is the number of 
substrate-type atoms present on the surface at the 
time step $i$.

Since our main goal is to obtain adsorption isotherms for different surface
structures of substrate atoms, we choose a simulation method which does not
yield information on the kinetics of nucleation and growth. The Kinetic
Monte Carlo (KMC) \cite{Rickvold,Schefler} technique could be applied with
this purpose. Studies in this direction are under way \cite{Ag-Au-dyn,
DMC-nuestro}.

\subsection{Algorithm employed for the calculation of energy differences}

One of the main advantages of the lattice model is its simplicity, since it
fixes the distances between the adsorption nodes, thus reducing the energy
values that the system can take to a discrete set. Furthermore, the
potentials used are short ranged, so that a very important simplifying
assumption can be made for obtaining $\Delta U.$ The point is to consider
the adsorption(desorption) of a particle at a node immersed in a certain
environment, as shown in Figure 1 . The adsorption site for
the particle is located in the central box, and the calculation of the
interactions is limited to a circle of radius $R.$ Then, the adsorption
energy for all the possible configurations of the environment of the central
atom can be calculated previous to the simulation. In this case, we employ
configurations which include first, second and third nearest neighbors,
given a total of $13$ sites, including the central atom (under
consideration). With this method all the adsorption energies of an
atom are tabulated, so that during the MC simulation the most expensive
numerical operations are reduced to the reconstruction of the number $I$
that characterizes the configuration surrounding the particle on the
adsorption node. Computationally speaking, $I$ is nothing but the index of
the array in which the energy is stored.

In all the structures generated, the plane of the adsorbates was located at
a vertical distance from the first substrate plane equivalent to the
distance between the substrate and one complete adsorbate monolayer. This
value was previously obtained by minimizing the energy as a function of that
distance.

\subsection{Simulated annealing}

Simulated annealing techniques have often been used to obtain minimal energy
structures or to solve ergodicity problems. A suitable way to implement them
is through the canonical Monte Carlo method at different temperatures. In
the present case, we used simulated annealing in order to obtain different
surface defects given by islands of various sizes and shapes. In all cases
the initial state involved a coverage degree of 0.1 substrate atoms distributed
at random.

The simulation was started at a very high initial temperature $T_o$, of the
order of 10$^4$ K, and the system was later cooled down following a
logarithmic law:

\begin{equation}
T_f=T_oK^{N_{cycles}}  \label{14}
\end{equation}

\noindent where T$_f$ is the final temperature, $N_{cycles}$ is the number of cooling
steps and $K$ is a positive constant lower than one (in this case, we use $%
K=0.9$). A certain number of MCS were run at each temperature and the
simulation stoped when $T_f$ was reached. The number of MCS employed at each
temperature varied from 20 to 655360, giving different kinds of
structures as we shall discuss later.

\subsection{Underpotential vs overpotential deposition}

The possibility that \textbf{upd }takes place can be quantified through the
so-called underpotential shift $\Delta \phi _{UPD},$ which is related to the
difference of the chemical potential of $M$ adsorbed on $S$ at a coverage
degree $\Theta $, say $\mu (M_\Theta /S)$, and the chemical potential of $M$
in the bulk phase, say $\mu (M/M)$, through the following equation:

\bigskip{} {\centering  }

{\ 
\begin{equation}
\Delta \phi _{UPD}=\frac{1}{ze_{0}}[\mu (M/M)-\mu (M_{\Theta }/S)]
\label{upd}
\end{equation}
}

\bigskip{}

\noindent where $z$ is the charge of the ion $M$ in the solution and $e_0$ is the
elemental charge. Since $\Delta \phi _{UPD}$ depends on the valence of the
ion being deposited, we prefer to use the excess of chemical potential $\Delta \mu $
as stability criteria for adsorbed monolayers which we define as:

\begin{equation}
\Delta \mu =\mu (M_\Theta /S)-\mu (M/M)  \label{delta_mu}
\end{equation}

According to equations \textbf{(4)} and \textbf{(5)}, negative values of $\Delta \mu$
indicate underpotential deposition, while positive ones predict overpotential
deposition.

\section{Results and discussion}

We have considered the following systems: Ag on Au(100); Au on Ag(100); Ag
on Pt(100); Pt on Ag(100); Au on Pt(100); Pt on Au(100), Pd on Au(100) and
Au on Pd(100). In the following discussion and for the sake of simplicity we
omit the Miller indices denoting the crystal surface, writing first the
adsobate and then the substrate, i.e. instead of Ag on Au(100) we
write Ag/Au.

\subsection{Energy tables}

In figure 1 we show 25 relevant configurations out of the 531441 which are 
possible for the environments that an atom can find on the surface.
The corresponding adsorption energies are summarized as illustrative examples 
in table I. 
Configurations 1-17 and 24 correspond to environments involving only
adsorbate atoms. Comparing configurations 1, 2, 3, 4 and 5, we can see the
influence of the first neighbors, which in all cases have a favorable effect
on the adsorption of the central atom. By comparing cases 6, 7, 8, 9 and
10, we can see the influence of second neighbors(in the absence of first
neighbors) and we can conclude that, except for Ag/Pt, they favor in all cases 
the adsorption of an atom at the central site.
Nevertheless, the influence of second neighbors seems to be the opposite
when first neighbors are present (see configurations 11 and 13 as
compared with configuration 2). This is due to the non-linear effects of the
many body interaction energies: the binding of the second neighbor to the
first one weakens the binding of the central atom to the first neighbor.
The direct influence of third neighbors is almost negligible (compare
configurations 1 and 16). On the other hand, the presence of a third
neighbor close to a first neighbor tends to be unfavorable for adsorption of the
central atom, for the same reasons stated in the previous case (compare
configurations 2 and 17).

Configurations 19 and 20 are meaningful when considering adsorption at step
sites. In this respect, observation of Table I indicates that when adsorbing 
on Pt steps Ag and Au atoms will avoid neighboring homoatoms. This is
not unexpected, since both Ag and Au exhibit an important compressive
surface stress when adsorbed on Pt. On the other hand, the adatoms of the
other systems will prefer to adsorb beside other adatoms.

Configurations 21, 22 and 23 represent the environment of the three kinds of
kink sites that an atom can find on the surface. Although no conclusion
can be drawn about a general trend, 
it can be stated that the energetic behavior is monotonic,
in the sense that the strength of the binding increases or decreases with
the coordination of the kink site for the three types of sites.

Comparing configurations 20 and 22 for all systems, it can be noticed that
for the systems Ag/Au, Ag/Pt and Au/Pt, adsorption onto kink sites(conf. 22)
is favored as compared with step decoration (conf. 20), whereas for the
systems Au/Ag, Pt/Ag, Pt/Au, Au/Pd and Pd/Au the growing of a
monodimentional phase at step sites should be preferred (conf. 20). This has
important consequences for the sequential filling of the different types of
deffects. In the case of the former family of systems, kink sites should be
filled before step decoration occurs, so that in the adsorption isotherm the
two processes should appear sequentially.  On the other hand, for the
second type of systems a sequential behavior should not be expected. We
shall return later to this phenomenon when analyzing the adsorption
isotherms. A similar trend can be noticed when comparing configurations 14 and
19, which correspond to the adsorption of an adatom close to a step of the
same (conf. 14) or a different (conf. 19) nature. For the systems Ag/Au, Ag/Pt,
Au/Pt and Au/Pd, configuration 19 is more stable than conf. 14, so that in
these cases, the adsorption of a new atom at the step of a substrate island
should be more favourable than adsorption on the edge of an adsorbate
island. The opposite trend is observed for the systems Au/Ag, Pt/Ag, Pt/Au
and Pd/Au. However the system Pd/Au is a borderline case, as the energy
difference between configurations 14 and 19 is of the order of 0.03 eV, that
is, close to kT at room temperature.

\subsection{Adsorption isotherms in defect free surfaces}

Adsorption isotherms were calculated for the different systems in the case
of defect free surfaces. For a fixed temperature and chemical potential a
simulation was performed and the average coverage degree was calculated for
the equilibrium state of the system. As this was repeated at different
chemical potentials, the isotherms obtained were like those represented 
in figure 2 for the case of the systems Ag/Au and Au/Ag.
The isotherms were
performed at a temperature of 300 K and we include two in each graphic:
A/S(100) and A/A(100), where A and S denote adsorbate and substrate 
respectively. Thus
comparison of the behavior of the heteroatomic A/S(100) system with that of
the pure metal A/A(100) system allows to determine the existence of
underpotential or overpotential deposition. As we can see, for each isotherm
there is an abrupt jump in the coverage degree as expected in the case of a
first order phase transition. As we discussed in a previous work \cite
{Ag-Au-thermo}, at higher temperatures the isotherms tend to become
smoother, approximating to the shape of a Langmuir isotherm.

According to equations \ (\ref{upd}) and (\ref{delta_mu}), the difference
between the chemical potentials $\Delta \mu$at which transition
occurs for the systems A/S(100) and A/A(100) is a measure of the
underpotential shift:

\begin{center}
$\Delta \phi _{UPD}=-\frac{\Delta \mu }{ze_{0}}\ $
\end{center}

In table 2, we summarize the approximate values for the chemical potential
at which transition occurs for each system. In table 3 we show the
excess of chemical potential for the systems A/S(100) as defined in equation
(\ref{delta_mu}), calculated as the difference of the values of the previous
table for the systems A/S(100) and A/A(100). Negative values predict
underpotential deposition, while positive ones predict overpotential
deposition. 
These results agree well with those obtained in previous works \cite
{upd-eam1,upd-eam2} by means of a thermodynamic cycle using the embedded 
atom method. According to these results, \textbf{upd} is predicted
for the systems  Ag/Au, Ag/Pt, Au/Pt and Au/Pd, and
overpotential deposition is expected for the remaining systems. Although this
prediction is in agreement with the experimental results for Ag/Au and
Ag/Pt, it does not agree with experimental finding of \textbf{upd }for
Pd/Au \cite{Kibler00}. It is also interesting to point out that this
system seems to go against the empirical rule that \textbf{upd} should occur 
when the work function of the metal being deposited is lower than that of the
substrate ( $\Phi _{Au}^{111}=$ 5.31eV; $\Phi _{Au}^{100}=$ 5.47eV; $\Phi
_{Au}^{110}=$ 5.37eV, $\Phi _{Pd}^{111}=$ 5.6eV \cite{Handbook}).
Underpotential deposition for this system also goes against the intuitive
expectation that it should occur when the surface energy $\sigma \ $ of the
substrate is larger than that of the adsorbate,($\sigma ^{Au}=0.094$ eV/\AA $%
^2$; $\ \sigma ^{Pd}=0.125$ eV/\AA $^2$, \cite{Daw-Baskes} ). First
principles calculations also yield a negative excess of binding energy\cite
{first-p-upd}, so that this system presents some remarkable singularities.
It must be emphasized, however, that Pd deposition occurs in the presence of 
strongly adsorbed chloride anions which result in a distorted-hexagon cloride on
top of the Pd monolayer\cite{Kibler00}.
Thus the presence of adsorbed anions may yield part of the energy excess
required for the existence of \textbf{upd.}

\subsection{Surface defects}

In order to study the influence of surface defects on the adsorption
isotherms, simulations were performed in the presence of substrate atoms in
the monolayer, with a coverage degree of 0.1. With the present simulation
parameters, such coverage degree corresponds to 1000 atoms as initial 
conditions. Different
surface defects were generated by employing various cooling rates in the
simulated annealing procedure described above. This was achieved using
different numbers of Monte Carlo steps $N_s$ at each temperature. Five
defective surface types were created, using $N_s$=20 for surface type 1, $%
N_s $=320 for surface type 2, $N_s$=5120 for surface type 3, $N_s$=81920 for
surface type 4 and $N_s$=655360 for surface type 5. Surface type 1
contains the largest quantity of islands and they are smaller than those of
other surfaces. The total number of islands decreases from surface types 1
to 5 and their size increase accordingly. This implies that the number of
kink sites and the number of step sites also decrease from surface type 1 to
5. In the following figures, these defective surface types are labelled
according to the average island size $<s_i>$ obtained, which were for example
7, 13, 48, 200 and 500 for the system Ag/Au.

\subsubsection{Adsorption isotherms}

As can be noticed in figure 2, adsorption isotherms on defect-free
surfaces show an abrupt change of coverage degree for a certain critical
chemical potential. This situation turns to be different in the presence of
surface defects for the systems Ag/Au, Ag/Pt, Au/Pd and Au/Pt (Figure 3). In
these cases the coverage degree starts to rise slowly at chemical potentials 
more negative than the critical chemical potential. This effect is
more important in the case of surfaces with smaller islands, which contain
more kink and step sites. Observation of the status of the surface,
discussed in detail below, shows that these are the places that are occupied
first. Relative coverage degrees\ $\Theta _k$ ($\Theta _s)$ 
for kink (step) sites may be defined as the number of occupied kink (step) 
sites divided by
the total number of kink (step) sites. Figures 4 and 5 show $\Theta _k,$ $%
\Theta _s$, and the total coverage degree $\Theta $ as a function of $\mu
_M $. The behavior of these isotherms indicates that kink sites are occupied
first, then, step sites, and finally, the rest of the surface. Close
inspection of these figures shows that the chemical potential values at which $%
\Theta _k$=0.5 are close to the corresponding energy values for adsorption at
kink sites (see configurations 21-23 in figure 1 and corresponding values in
Table 1). Something similar occurs with the $\mu _M$ values at which $\Theta
_s$=0.5, which are close to the energy values of configuration 20 in figure
1 (see also entry 20 in Table 1). However, it must be noticed that an abrupt
jump is not observed in the $\Theta _k$ and $\Theta _s$ \ isotherms, as
expected for the 0 and 1 dimensional systems.

Unlike the systems considered in figure 3, the systems Au/Ag,
Pt/Ag, Pd/Au and Pt/Au do not present an appreciable widening of the
adsorption isotherms with increasing number of surface defects (Figure 6).
Figures 7 and 8 show that the partial coverages $\Theta _k$ and $\Theta _s$
for this type of systems do not exhibit a clearcut trend as 
the systems depicted in figures 4 and 5. This can be understood through
comparative analysis of configuration 20 against configurations 21-23. It is
clear that these systems do not prefer adsorption at steps to adsorption at kink
sites, in contrast to the prediction for the systems presented in figure 3.
Thus kink sites are not filled before step decoration, but both processes
take place simultaneously. Furthermore, as pointed out above,
configuration 14 yields lower energy values than configuration 19 does, so that
atom deposition at substrate island steps is delayed with respect to
deposition at an island of the same nature. It is also remarkable that
within the present model all these systems exhibit a positive excess of
binding energy, and that the binding energy of the adsorbate is larger than, or
similar to that of the substrate (See table 4).

\subsubsection{Morphology}

It is very illustrative to analyze the final state of the surface of some
typical systems at different chemical potentials. In figure 9, we show
frames corresponding to the final state of the Ag/Pt system at 6 different
chemical potentials. In principle, the islands are not fixed and all atoms
are allowed to diffuse, however, there appears to be no movement of the
substrate type atoms, so that the islands remain unchanged. 
The adsorbate atoms
occupy the free sites with the preferences described above, that is, first
filling the kink sites, then the steps, and finally the rest of the system. A
similar behavior is observed for the systems Ag/Au and Au/Pt. The situation
is strikingly different for the systems Au/Ag, Pt/Ag, Pd/Au and Pt/Au, in
all of which, adsorbates come into the islands forming an alloy and these
tend to disintegrate. This is illustrated in figure 10 for Pt
deposition on Ag(100). At small coverage degrees the adatoms start to get
already embedded into the islands, and even form small nucleii. This process
continues and upon completion of the monolayer the substrate islands have
incorporated an important number of adatoms. This phenomenon also occurs for
larger islands, as can be observed in figure 11, where this process is
illustrated for Pt deposition on an Ag surface exhibiting a big 1000-atom 
island. In this case, and for larger coverage degrees, a higher concentration
of adatoms results on the edge of the islands due to a slower equilibration
of the system. The system Au/Pd is an intermediate case.

A detailed view of the neighborhood of an island is presented in figures 12
and 13 for the systems Au/Pt and Pt/Au, showing the features 
mentioned in the previous discussion.

While the simplicity of the present model (not considering surface
reconstruction, anion adsorption and specific kinetic features) does not allow
a quantitative comparison with experiments, qualitative predictions can be made
concerning surface alloy formation in the presence of islands.
In table 5 we present the predictions for the systems simulated in this work
along with some results of experimental observation. In connection with 
the present results, it is worth mentioning the ones obtained by Kolb and 
coworkers for Pd and Pt deposition on Au(100) \cite{Kibler00, Waibel02}. 
For the former
system, the authors proposed that alloying upon Pd deposition should proceed
involving Au atoms from islands and step edges. The present results strongly
support this explanation on thermodynamic grounds, based on the energetics
of the Pd/Au system. 
Interesting results have also been obtained for the second system, Pt/Au. 
Waibel et al \cite{Waibel02} have studied Pt
deposition on Au(100), finding that nucleation of Pt starts mainly at
deffects like step edges for low deposition rates. On the other hand, at
high deposition rates some nucleii also appear on the terraces at random
sites. Figure 7 of reference \cite{Waibel02} shows that over Pt deposition
the island shape becomes progressively blurred as Pt is deposited, and it is 
very stable. According to the present results, at low deposition rates Pt atoms
could be incorporating into the islands, yielding the enhanced stability
observed. Due to its high binding energy, Pt is expected to present
3-D growth, as pointed out by Waibel et al, but this feature is not
considered in the present model.

\subsubsection{First neighbor site occupation}

In order to get a more complete picture of phase growth in the presence of
substrate islands, the average number of first neighbors of a given atom
type ($M$ or $S$) surrounding an occupied site (with $M$ or $S$) was also
analyzed. $n_{A}^{B}$ will denote the average number of neighbors of 
$B$ type atoms surrounding an $A$ type atom. Let us consider first the
system Au/Pt, depicted on the left side in the lower part of figure 4. It
can be seen that the average number of Pt atoms surrounding other Pt atoms
as nearest neighbors , $n_{Pt}^{Pt}$, remains constant, at a value somewhat
lower than four. This shows that the coordination of Pt atoms on the islands
does not change upon Au deposition onto Pt, in agreement with the fact that Pt
atoms remain almost immobile. As expected, simulations with larger Pt
islands yield $n_{Pt}^{Pt}$ closer to 4, while the reverse occurs with
simulations with smaller islands. The number of Au atoms around Pt, 
$n_{Pt}^{Au},$ rises from zero at chemical potentials where $Au$ is
absent, to a number less than one, when the monolayer is complete. This is
due to the fact that only Pt atoms on the edge of the islands are in
contact with adsorbate atoms. Two steps can be noticed in the 
$n_{Pt}^{Au}-\mu $ curve. The first one occurs when the kink sites of the Pt
islands become occupied by Au atoms, and the second one when the edges of the
islands (step sites) become decorated with Au. By observing the average number of 
$Pt$ atoms around $Au$, $n_{Au}^{Pt},$it can be noticed that $%
n_{Au}^{Pt}\approx 2$ in regions where only kink sites are covered, close to
one in regions where step sites are covered, and very small upon completion of
the $Au$  monolayer. Some isolated cases with $n_{Au}^{Pt}=3$ can be
observed for low coverage degrees. This is due to the fact that individual
atoms of adsorbate manage to incorporate to the edge of the island, but
are unable to penetrate further. The number of $Au$ atoms around $Au$ atoms, 
$n_{Au}^{Au},$ is zero for $\Theta _{Au}\longrightarrow 0$ and in regions
where only kink sites are occupied. $n_{Au}^{Au}$ then rises to a number
between one and two in regions where steps sites are occupied and becomes
close to four for monolayer completion. The general features of the system
Ag/Pt, depicted at the bottom right part of figure 4 are similar.
This picture of sequential filling of kink and step sites is also
reflected by the partial isotherms at the top of figure 4: the
coverage degree of steps sites starts to rise only when $\Theta _{k}\approx
1$.

Figure 5 shows the same analysis for the systems Au/Pd and Ag/Au. In the
lower part of the right hand side, it can be observed for the system Ag/Au
that the general trends are similar to those of the two previous systems,
but the curves $n_{adsorbate}^{substrate}$ ($n_{Ag}^{Au}$)do not present two
clearly differentiated steps as before. That is, the regions with occupation
of kinks and steps are not clearly diferentiated. The same can be stated 
for the system Au/Pd (left hand side in figure 5). In this
figure, the curve $n_{Au}^{Pd}$ presents an even more remarkable behavior.
At low coverage degrees, the average number of Pd atoms around Au is close
to four, indicating that they came inside the existing substrate islands
and are completely surrounded by Pd atoms. Thus, although $n_{Pd}^{Pd}$
remains fairly constant denoting an important island stability, some small
degree of alloying occurs. Thus the Au/Pd system is a borderline case,
where some alloying may occur but islands still remain stable.
The non sequential filling of kink and step sites can also be inferred
from the partial isotherms in the upper part of figure 5, where it can be
observed that $\Theta _{s}$ starts to rise when the decoration of kink sites
is still incomplete.

The lower part of figures 7 and 8 show $n_{A}^{B}$ vs $\mu $ plots for
Pd/Au, Pt/Ag, Pt/Au and Au/Ag. It is evident that in all cases the number of
substrate atoms around substrate atoms ($n_{Au}^{Au}$ y $n_{Ag}^{Ag}$) 
decreases for increasing coverage
degree, denoting that the islands do not remain unaltered in the presence of
the adsorbate. In the case of Pd adsorption on Au(100) and Au adsorption on
Ag(100), it can be noticed that at low coverage degrees the adsorbate atoms
are completely embedded in the islands ($n_{Pd}^{Au}=4$ for Pd/Au and $%
n_{Au}^{Ag}=4$ for Au/Ag). This effects somehow occur for Pt/Ag and Pt/Au,
but to a lesser extent. In the case of Pt adsorption, the number of substrate
atoms around Pt is smaller than four because clustering of Pt atoms occur
within the substrate islands, even at low Pt coverage degrees. This can be 
observed in the frames 2-4 of figure 10.

The behavior of $n_{substrate}^{substrate}$ at high $\mu $ for the
systems Pt/Ag, Pt/Au and Au/Ag presents one very interesting feature: 
$n_{substrate}^{substrate}$ remains approximately
constant over a certain $\mu $ region, when surface coverage by the
adsorbate is close to completion. This can be observed, for instance, at the
bottom of the right hand of figure 8, where $n_{Ag}^{Ag}$ remains nearly
constant above $\mu =$ -3.9 eV. As a counterpart, $n_{Ag}^{Au}$ also remains
constant in this region. This means that the surface presents patches of
constant composition that remain stable. In other words, in those places
where the islands were originally located, patches appear which exhibit a
surface larger than that of the original island. This analysis complements the
discussion in the last part of the previous section.

\section{Conclusions}

\begin{enumerate}

\item  The systems Ag/Au(100), Ag/Pt(100), Au/Pt(100) and Au/Pd(100) present
a positive excess of binding energies (negative excess of chemical
potential) as compared with the homoepitaxial growth of adsorbate type atoms,
indicating that in these systems underpotential deposition is expected.

\item  On the other hand, for the Au/Ag(100), Pt/Ag(100), Pt/Au(100), and
Pd/Au(100) systems, the monolayer adsorption is more favorable on substrates
of the same nature than on the substrates considered.

\item  In order to emulate surface defects, simulations were also performed 
in the presence of surface islands made of the same metal as  the substrate. 
For the family of systems mentioned in item 1, the islands remained
almost unchanged, being decorated by the adatoms before completion of the
monolayer. 
In the case of the systems Au/Pt(100) and Ag/Pt(100) the
adsorbate atoms filled in a clear sequence first the kink and then the step sites.
These processes are somewhat closer in the case of Ag/Au(100) and very close in
Au/Pd(100).  For
the family of systems considered in item 2, the substrate islands showed
disgregation in order to form 2-D alloys with the adsorbate atoms and there is
no differentiation in the filling of kink, step or terrace sites. 

\item The system Au/Pd(100) presents a borderline behavior, as a small 
quantity of Au is embedded into the Pd islands without altering 
their structure.
Thus it could be suggested that, except for the systems containig Au and Pd, the
stability of the substrate islands 
upon deposition of a foreing metal is mainly determined by the 
difference of the cohesive energies. However, this is a matter that 
requires further study to make a more conclusive statement.

\end{enumerate}


\section{Acknowledgements}

We thank CONICET, SeCyt-UNC, Agencia C\'{o}rdoba Ciencia, Program BID
1201/OC-AR PICT $N^{o}$ 06-04505.

We thank Mario Del Popolo for stimulating discussion and ideas.

Language assistance by Karina Plasencia is gratefully acknowledged.


\newpage


\section{Tables}

\textbf{Table 1:} Energy differences (in eV) associated with the deposition of
an atom in the environments represented in figure 3 for the systems
considered in this work.

\vspace{1cm}

\begin{tabular}{|l|l|l|l|l|l|l|l|l|}
\hline
$conf.$ & $Ag/Au$ & $Au/Ag$ & $Ag/Pt$ & $Pt/Ag$ & $Au/Pt$ & $Pt/Au$ & $Au/Pd$
& $Pd/Au$ \\ \hline
$1$ & -2.580 & -3.106 & -3.127 & -4.224 & -3.672 & -4.228 & -3.283 & -3.117
\\ \hline
$2$ & -2.866 & -3.566 & -3.340 & -4.828 & -4.160 & -4.847 & -3.901 & -3.502
\\ \hline
$3$ & -3.126 & -3.987 & -3.521 & -5.397 & -4.585 & -5.434 & -4.444 & -3.862
\\ \hline
$4$ & -3.366 & -4.380 & -3.681 & -5.938 & -4.965 & -5.986 & -4.453 & -3.863
\\ \hline
$5$ & -3.591 & -4.744 & -3.825 & -6.455 & -5.300 & -6.508 & -5.358 & -4.535
\\ \hline
$6$ & -2.582 & -3.142 & -3.122 & -4.267 & -3.705 & -4.246 & -3.335 & -3.137
\\ \hline
$7$ & -2.585 & -3.177 & -3.116 & -4.310 & -3.738 & -4.254 & -3.388 & -3.157
\\ \hline
$8$ & -2.585 & -3.177 & -3.116 & -4.310 & -3.738 & -4.264 & -3.388 & -3.157
\\ \hline
$9$ & -2.588 & -3.212 & -3.110 & -4.353 & -3.770 & -4.282 & -3.439 & -3.178
\\ \hline
$10$ & -2.590 & -3.247 & -3.105 & -4.396 & -3.802 & -4.300 & -3.490 & -3.198
\\ \hline
$11$ & -2.842 & -3.563 & -3.303 & -4.837 & -4.130 & -4.833 & -3.878 & -3.497
\\ \hline
$12$ & -2.868 & -3.599 & -3.332 & -4.869 & -4.188 & -4.863 & -3.948 & -3.521
\\ \hline
$13$ & -2.825 & -3.567 & -3.275 & -4.852 & -4.112 & -4.817 & -3.863 & -3.500
\\ \hline
$14$ & -2.810 & -3.541 & -3.259 & -4.826 & -4.071 & -4.785 & -3.815 & -3.485
\\ \hline
$15$ & -3.076 & -3.947 & -3.455 & -5.370 & -4.495 & -5.388 & -4.347 & -3.833
\\ \hline
$16$ & -2.580 & -3.106 & -3.126 & -4.224 & -3.671 & -4.227 & -3.282 & -3.116
\\ \hline
$17$ & -2.840 & -3.529 & -3.309 & -4.792 & -4.103 & -4.814 & -3.833 & -3.477
\\ \hline
$18$ & -3.002 & -3.470 & -3.692 & -4.634 & -4.336 & -4.708 & -3.905 & -3.497
\\ \hline
$19$ & -2.917 & -3.473 & -3.564 & -4.651 & -5.139 & -4.640 & -3.885 & -3.450
\\ \hline
$20$ & -3.149 & -3.859 & -3.174 & -5.185 & -4.626 & -5.193 & -4.382 & -3.777
\\ \hline
$21$ & -3.250 & -3.760 & -4.010 & -4.996 & -4.758 & -5.048 & -4.351 & -3.758
\\ \hline
$22$ & -3.230 & -3.770 & -3.976 & -5.009 & -4.747 & -5.029 & -4.363 & -3.747
\\ \hline
$23$ & -3.211 & -3.780 & -3.942 & -5.022 & -4.736 & -5.010 & -4.374 & -3.736
\\ \hline
$24$ & -3.358 & -3.498 & -3.534 & -6.257 & -4.808 & -6.176 & -4.796 & -4.384
\\ \hline
$25$ & -3.731 & -4.287 & -4.621 & -5.661 & -5.505 & -5.635 & -5.176 & -4.223
\\ \hline
\end{tabular}

\newpage

\textbf{Table 2:} Chemical potential \ $\mu $ in eV at which the step is
observed in the adsorption isotherms for the different systems.

\vspace{1cm}

\begin{tabular}{|l|l|l|l|l|}
\hline
$Substrate\backslash Adsorbate$ & $Ag$ & $Au$ & $Pt$ & $Pd$ \\ \hline
$Ag$ & -2.83 & -3.87 & -5.30 & ---- \\ \hline
$Au$ & -3.00 & -3.95 & -5.29 & -3.78 \\ \hline
$Pt$ & -3.38 & -4.37 & -5.83 & ---- \\ \hline
$Pd$ & ---- & -4.21 & ---- & -3.92 \\ \hline
\end{tabular}

\vspace{1cm}

\textbf{Table 3:} Excess of chemical potential in eV, as calculated from the
adsorption isotherms according to equation (\ref{delta_mu}).

\vspace{1cm}

\begin{tabular}{|l|l|l|l|l|}
\hline
$substrate\backslash adsorbate$ & $Ag$ & $Au$ & $Pt$ & $Pd$ \\ \hline
$Ag$ & 0.00 & 0.08 & 0.53 & ---- \\ \hline
$Au$ & -0.17 & 0.00 & 0.54 & 0.14 \\ \hline
$Pt$ & -0.55 & -0.42 & 0.00 & ---- \\ \hline
$Pd$ & ---- & -0.26 & ---- & 0.00 \\ \hline
\end{tabular}

\vspace{1cm}

\textbf{Table 4:} Cohesive energies of the bulk metal $E_{coh}^{M(s)}$ \cite
{Daw-Baskes}.

\vspace{1cm}

\begin{tabular}{|l|l|l|l|l|}
\hline
Metal & Ag & Pd & Au & Pt \\ \hline
$E_{coh}^{M(s)}$ & -2.85 & -3.91 & -3.93 & -5.77 \\ \hline
\end{tabular}

\newpage

\textbf{Table 5:} Prediction of surface alloy in the presence of islands
for systems simulated in this work, and experimental observations.

\vspace{1cm}

\begin{tabular}{|l|l|l|}
\hline
System & Surface alloy prediction & Experimental observation \\ \hline
Ag/Au(100) & No & No \\ \hline
Ag/Pt(100) & No & No \\ \hline
Au/Pd(100) & Sligth alloying & Not available \\ \hline
Au/Pt(100) & No & Not Available \\ \hline
Au/Ag(100) & Yes & Nor Available \\ \hline
Pt/Ag(100) & Yes & Not Available \\ \hline
Pd/Au(100) & Yes & Surface alloying ref \cite{Kibler00} \\ \hline
Pt/Au(100) & Yes & Au islands became stable agains dissolution ref \cite
{Waibel02} \\ \hline
\end{tabular}

\newpage


\section{Figure Captions}

\textbf{Figure 1:} Some of the possible configurations associated with the
adsorption of the central atom in different environments. The energies
associated with all posible configurations were tabulated previous to the
simulation.
The energy values corresponding to these examples are shown in table 1. 
\vspace{0.5cm}

\textbf{Figure 2:} Adsorption isotherms, plotted as coverage degree as a
function of chemical potential. \textbf{a)} Deposition of one monolayer of
Ag onto a defect-free Au(100) surface and one monolayer of Ag onto a defect-free
Ag(100) surface. \textbf{b)} Deposition of one monolayer of Au onto a 
defect-free Ag(100) surface and one monolayer of Au onto a defect-free Au(100)
surface. 
\vspace{0.5cm}

\textbf{Figure 3:} Adsorption isotherms, plotted as coverage degree as a
function of chemical potential, for the deposition of Ag on Au(100), Ag on
Pt(100), Au on Pd(100) and Au on Pt(100) in presence of surface
defects. The numbers indicate the average size of islands of 
substrate-type atoms present in the monolayer as initial state. 
\vspace{0.5cm}

\textbf{Figure 4:} Adsorption isotherms for the deposition of Au on Pt(100)
and Ag on Pt(100) in the presence of surface defects. Coverage degree of the
monolayer, the step sites and the kink sites as a function of chemical
potential (up). Average number of neighbors of each species for each kind of
atom (down). 
\vspace{0.5cm}

\textbf{Figure 5:} Adsorption isotherms for the deposition of Au on Pd(100)
and Ag on Au(100) in the presence of surface defects. Coverage degree of the
monolayer, the step sites and the kink sites as a function of chemical
potential (up). Average number of neighbors of each species for each kind of
atom (down). 
\vspace{0.5cm}

\textbf{Figure 6:} Adsorption isotherms plotted as coverage degree as a
function of chemical potential for the deposition of Au on a Ag(100), Pt on
a Ag(100), Pd on a Au(100) and Pt on a Au(100) in the presence of surface
defects. The numbers indicate the average size of islands of substrate-type atoms
present in the monolayer as initial state. 
\vspace{0.5cm}

\textbf{Figure 7:} Adsorption isotherms for the deposition of Pd on Au(100)
and Pt on Ag(100) in thepresence of surface defects. Coverage degree of the
monolayer, the step sites and the kink sites as a function of chemical
potential (up). Average number of neighbors of each species for each kind of
atom (down). 
\vspace{0.5cm}

\textbf{Figure 8:} Adsorption isotherms for the deposition of Pt on Au(100)
and Au on Ag(100) in the presence of surface defects. Coverage degree of the
monolayer, the step sites and the kink sites as a function of chemical
potential (up). Average number of neighbors of each species for each kind of
atom (down). 
\vspace{0.5cm}

\textbf{Figure 9:} Snapshots of the final state of the surface at six
different chemical potentials (-4.27 eV, -3.97 eV, -3.66 eV, -3.44 eV, -3.40
eV and -3.06 eV) for the Ag on Pt(100) simulation. Average island size: $%
<s>=48$. 
\vspace{0.5cm}

\textbf{Figure 10:} Snapshots of the final state of the surface at six
different chemical potentials (-5.74 eV, -5.53 eV, -5.41 eV, -5.32 eV, -5.30
eV and -5.21 eV) for the Pt on Ag(100) simulation. Average island size: $%
<s>=53$. 
\vspace{0.5cm}

\textbf{Figure 11:} Snapshots of the final state of the surface at six
different chemical potentials (-5.74 eV, -5.53 eV, -5.41 eV, -5.32 eV, -5.30
eV and -5.21 eV) for the Pt on Ag(100) simulation. Average island size: $%
<s>=1000.$ (only one island). 
\vspace{0.5cm}

\textbf{Figure 12:} Final state of the surface at four different chemical
potentials (-5.10 eV, -4.61 eV, -4.50 eV and -4.39 eV) for the Au on Pt(100)
simulation with defects. One Pt island. White circles: gold atoms. Black
circles: Pt atoms. 
\vspace{0.5cm}

\textbf{Figure 13:} Final state of the surface at four diferent chemical
potentials (-5.65 eV, -5.30 eV, -5.29 eV and -5.26 eV) for the Pt on Au(100)
simulation with defects. White circles: gold atoms. Black circles: Pt atoms. %
\vspace{0.5cm}


\begin{thebibliography}{99}

\bibitem{Kolb_book}  D.M. Kolb, in: H. Gerischer, C.W. Tobias(Eds.).
\textit{Advances in Electrochemistry and Electrochemical Engineering}. vol 11, Wiley,
New York \textbf{1978}, page 125.

\bibitem{Kolb}  D.M. Kolb, M. Przasnyski and H. Gerischer, J. Electroanal.
Chem. \textbf{1974}, \textit{54}, 25.

\bibitem{Trasatti}  S. Trasatti, Zeits. f\"{u}r Phys. Chem. NF  \textbf{1975},
 \textit{98}, 75.

\bibitem{Blum}  L. Blum, D. A. Huckaby and M. Legault, Electrochim. Acta 
\textbf{1996}, \textit{41}, 2201.

\bibitem{Lorenz_2}  K. Jutttner, G. Staikov, W.J. Lorenz, E. Schmidt, J.
Electroanal. Chem. \textbf{1977}, \textit{80}, 67.

\bibitem{Leiva_UPD_1}  E. P. M. Leiva, in ``Current Topics in
Electrochemistry'', Council of Scientific Information,
Trivandrum, India \textbf{1993}, \textit{2}, 269.

\bibitem{Leiva_UPD_2}  E. P. M. Leiva, Electrochim. Acta \textbf{1996}, 
\textit{41}, 2185.

\bibitem{Lorenz_Libro}  E. Budevski, G. Staikov and W. J. Lorenz, \textit{\
Electrochemical Phase Formation and Growth },VCH Weinheim \textbf{1996}.

\bibitem{Daw-Baskes}  S. M. Foiles, M. I. Baskes and M. S. Daw, Phys. Rev. B 
\textbf{1986}, \textit{33},7983.

\bibitem{Pb/Ag}  M. G. Samant, M. F. Toney, G. L. Borges, L. Blum and O. R.
Melroy, J. Phys. Chem \textbf{1998}, \textit{92}, 220.

\bibitem{Ag-Au-thermo}  M.C. Gim\'{e}nez, M.G. Del P\'{o}polo and E.P.M.
Leiva, Electrochimica Acta \textbf{1999}, \textit{45},699.

\bibitem{first-p-upd}  C. G. S\'{a}nchez, E.P.M. Leiva and J. Kohanoff,
Langmuir, C. S\'{a}nchez, E.P.M. Leiva and J. Kohanoff, Langmuir \textbf{2001}, 
\textit{17}, 2219.

\bibitem{tablas_eam_1}  E. Clementi, C. Roetti, At. Data Nucl. Data Tables
\textbf{1974}, \textit{14}, 177.

\bibitem{tablas_eam_2}  A. D. McLean, R. S. McLean, At. Data Nucl. Data
Tables  \textbf{1981}, \textit{26}, 197.

\bibitem{Allen}  M. P. Allen and D. J. Tildesley, Computer Simulation of
Liquids, Oxford University Press \textbf{1987}.

\bibitem{Rickvold}  G. Brown, P. A. Rickvold, M. A. Novotny and A.
Wieckowski, Colloid and Surfaces A, in Press, or cond-mat/9703209.

\bibitem{Schefler}  P. Ruggerone, C. Ratsch and M. Scheffler, \textit{Growth of
Ultrathin Epitaxial Layers}, Vol. 8, Eds. D. A. King and D. P. Woodruff,
Elsevier Science, Amsterdam \textbf{1997}.

\bibitem{Ag-Au-dyn}  M.C. Gim\'{e}nez, M.G. Del P\'{o}polo, E.P.M. Leiva,
S.G. Garc\'{\i}a, D.R. Salinas C.E. Mayer and W.J. Lorenz, J. Electrochem.
Soc. \textbf{2002}, \textit{149}, E109.

\bibitem{DMC-nuestro}  M.C. Gim\'{e}nez, M.G. Del P\'{o}polo, E.P.M. Leiva,
Langmuir \textbf{2002}, \textit{18}, 9087-9094.


\bibitem{upd-eam1}  C. G. S\'{a}nchez, M. G. Del P\'{o}polo and E.P.M.
Leiva, Surface Science \textbf{421}, \textit{59} (1999).

\bibitem{upd-eam2}  M. I. Rojas, C. G. S\'{a}nchez, M. G. Del P\'{o}polo and
E.P.M. Leiva, Surface Science \textbf{2000}, \textit{453}, 225.

\bibitem{Handbook}  CRC Handbook of Chemistry and Physics, 80$^{th}$
Edition, D.R. Lide, Ed., Boca Rat\'{o}n \textbf{2000}.

\bibitem{Kibler00}  L.A. Kibler, M. Kleinert and D.M. Kolb, Surface Science
\textbf{2000}, \textit{461}, 155.

\bibitem{Waibel02}  H.-F. Waibel, M. Kleinert, L.A. Kibler and D.M. Kolb,
Electrochim. Acta \textbf{2002}, \textit{47}, 1461.
\end{thebibliography}
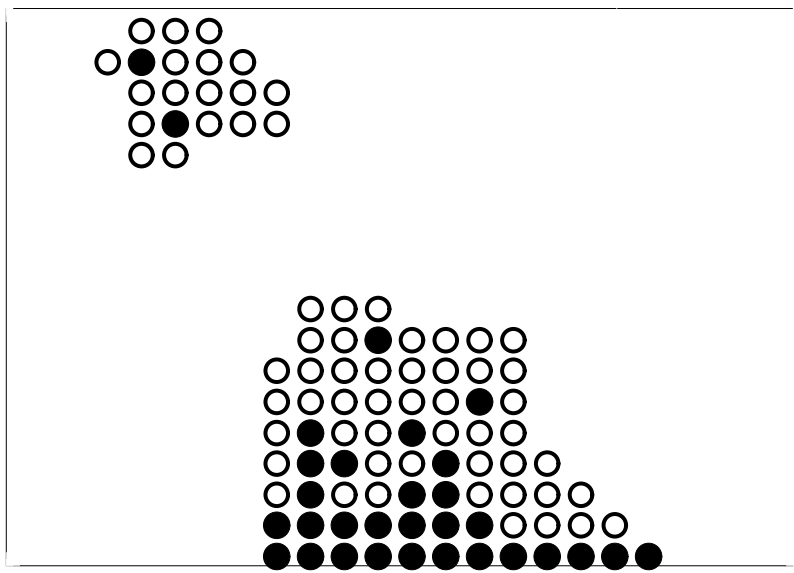
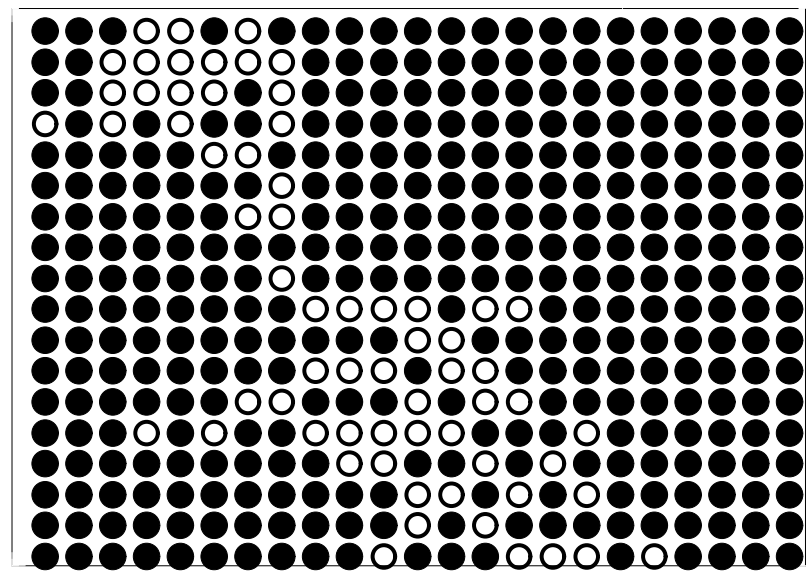
\end{document}